\documentclass[a4paper,fleqn,usenatbib,useAMS]{mnras}

\usepackage{graphicx}   % Including figure files
\usepackage{amsmath}    % Advanced maths commands
\usepackage{amssymb}    % Extra maths symbols
\usepackage{multicol}   % Multi-column entries in tables
\usepackage{bm}         % Bold maths symbols, including upright Greek
\usepackage{pdflscape}  % Landscape pages

\def\la{\;
\raise0.3ex\hbox{$<$\kern-0.75em\raise-1.1ex\hbox{$\sim$}}\; }
\def\ga{\;
\raise0.3ex\hbox{$>$\kern-0.75em\raise-1.1ex\hbox{$\sim$}}\; }

\newcommand{\dmm}{$\Delta\mu/\mu$}

\newcommand{\kms}{km~s$^{-1}$}
\newcommand{\etal}{{et al.}}

\newcommand{\CI}{[C {\sc i}]}
\newcommand{\CII}{[C {\sc ii}]}

\usepackage[T1]{fontenc}
\usepackage{ae,aecompl}

\title[Testing the WEP in the LMC/SMC]{\textit{Testing the weak equivalence principle by
differential measurements of fundamental constants in the
Magellanic Clouds }}

\author[S. A. Levshakov et al.]{
S. A. Levshakov$^{1,2}$\thanks{E-mail: lev@astro.ioffe.ru},
K.-W. Ng$^{3,4}$,
C. Henkel$^{5,6}$,
B. Mookerjea$^{7}$,
I. I. Agafonova$^{2}$,
\newauthor
S.-Y. Liu$^4$
and W.-H. Wang$^4$
\vspace*{8pt}
\\
$^{1}$Ioffe Physical-Technical Institute, 194021 St.~Petersburg, Russia\\
$^{2}$Electrotechnical University ``LETI'', 197376 St.~Petersburg, Russia \\
$^{3}$Institute of Physics, Academia Sinica, Taipei 11529, Taiwan\\
$^{4}$Institute of Astronomy and Astrophysics, Academia Sinica, Taipei 11529, Taiwan\\
$^{5}$Max Planck Institut f\"ur Radioastronomie, Auf dem H\"ugel 69, 53121 Bonn, Germany\\
$^{6}$Astron. Dept., King Abdulaziz University, PO Box 80203, 21589 Jeddah, Saudi Arabia\\
$^{7}$Tata Institute of Fundamental Research, Homi Bhabha Road, 400005 Mumbai, India }

\date{Accepted XXX. Received YYY; in original form ZZZ}

\pubyear{2019}

\begin{document}
\label{firstpage}
\pagerange{\pageref{firstpage}--\pageref{lastpage}}
\maketitle

\begin{abstract}
Non-standard fields are assumed to be responsible for phenomena attributed to
dark energy and dark matter. Being coupled to ordinary matter, these fields 
modify the masses and/or charges of the elementary particles, thereby 
violating the Weak Equivalence Principle. Thus, values of fundamental constants such as  
the proton-to-electron mass ratio, $\mu$, and/or the fine structure constant, $\alpha$, 
measured in different environment conditions can be used as probes for this coupling.  
Here we perform differential measurements of $F = \mu \alpha^2$ to test a non-standard
coupling in the Magellanic Clouds~-- dwarf galaxies where the overall mass budget is 
dominated by dark matter. The analysis is based on [C\,{\sc i}] and CO lines 
observed with the {\it Herschel Space Observatory}. Since these lines have different 
sensitivities to changes in $\mu$ and $\alpha$, the combined $\alpha$ and $\mu$ variations 
can be evaluated through the radial velocity offsets, $\Delta V$, between the CO and 
[C\,{\sc i}] lines. Averaging over nine positions in the Magellanic Clouds, we obtain 
$\langle \Delta V \rangle = -0.02\pm0.07$ \kms, leading to 
$|\Delta F/F| < 2\times10^{-7}$ ($1\sigma$),
where $\Delta F/F = (F_{\rm obs}-F_{\rm lab})/F_{\rm lab}$.
However, for one position observed with five times higher spectral resolution we find  
$\Delta V = -0.05\pm0.02$ \kms, resulting in $\Delta F/F = (-1.7\pm0.7)\times10^{-7}$.
Whether this offset is due to changes in the fundamental constants, due to
chemical segregation in the emitting gas or merely due to Doppler noise  
requires further investigations.
\end{abstract}


% BODY OF PAPER 

\section{Introduction}
\label{Sec1}

The weak equivalence principle (WEP) is one of the basic postulates of 
Einstein's general relativity (GR).
It assumes the equality of the inertial and gravitational mass, or, in terms of the field theory,
the universal and minimal coupling of all matter fields to a single metric. 
Up to now, GR (and, hence, WEP) successfully withstands all possible tests including those
performed under extreme gravity (e.g., Archibald \etal\ 2018; Will 2014).
However, it is widely believed that $\sim 95$\% of
the energy density of the Universe is concentrated in the so called `{\it dark sector}'
comprising the non-baryonic dark matter (DM, $\sim 26$\%) 
used to explain the CMB spectrum,
the formation of the large-scale structure and the observed discrepancy between visible and dynamical 
masses in galaxies and clusters,
and the dark energy (DE, $\sim 69$\%) 
responsible for the cosmic acceleration at low redshifts $(z < 1)$.
Both these substances cannot be understood within the framework of GR 
and the Standard Model of particle physics. 

To explain the cosmic acceleration either new scalar fields are explicitly introduced, 
or the Einstein field equations are modified what effectively is equivalent to the introduction 
of additional fields (Brax 2018; Nojiri \etal\ 2017; Joyce \etal\ 2016).
Some models allow these fields to couple to baryonic matter.
As for the DM, the paradigm for most favorable candidates shifts now from weakly interacting 
massive ($m \sim$ GeV) particles (WIMPs) to light or ultra-light 
($m \sim 10^{-3}$ eV down to $\sim 10^{-22}$ eV)
axion-like particles (ALPs) (Hui \etal\ 2017; Berezhiani \& Khoury 2015; Feng 2010). 
The main difference between the WIMPs and ALPs is that WIMPs are collisionless and interact only 
via gravitation, whereas ALPs are self-interacting, behave under special conditions as a quantum liquid 
and can fundamentally couple to ordinary matter (Irastorza \& Redondo 2018).

Additional coupling~-- being dependent on space, time and/or chemical composition~-- would
break the condition of universal coupling to the metric, thereby violating the WEP. 
Since detection of any violation of the WEP would manifest the presence of an unknown interaction and, 
hence, a new physics beyond the Standard Model, 
various laboratory and satellite experiments were carried out aimed 
at testing the WEP with highest precision possible. 
Up to now, nowhere any signal over the background was found
(Berg\'e \etal\ 2018; Antoniou \& Perivolaropoulos 2017; 
Rider \etal\ 2016; Brax \& Davis 2016; Li \etal\ 2016; Hamilton \etal\ 2015; Wagner \etal\ 2012).

One thing to notice is that if the anticipated new fields are so light as considered in theory, then
their gradients (fifth force) would be detectable only at scales of an order of (inter)galactic
distances. Another reason for null results could be a screening (damping) of these fields in the
environments where the experiments were performed. 
Several screening mechanisms were proposed and widely discussed in the literature, e.g.
a popular chameleon field model suggested by Khoury \& Weltman (2004), 
when coupling strength depends on the environmental matter density, or
a concept of emergent gravity recently developed by Verlinde (2017), 
where additional coupling is modulated by surface mass density of baryons. 
However, non of these screening models has yet been verified experimentally.

In such a case it seems natural to search
for a new field-baryon coupling just in places where the presence of non-standard fields could
be supposed from the phenomenology. 
In this respect the position of the Sun in the Milky Way (MW) is
not very favorable: different approaches to modeling the MW rotational curve come
to the same conclusion that the total matter within the solar
circle is dominated by baryons with 
$\rho_{\rm b}/\rho_{\scriptscriptstyle{\rm DM}} \sim 10$, 
where $\rho_{\rm b}$ is the baryon density and 
$\rho_{\scriptscriptstyle{\rm DM}}$ is the density of DM  
(McGaugh 2018; McMillan 2017; Iocco \etal\ 2015; Sofue \etal\ 2009).
On the other hand, the effects attributed to DM 
become prominent either at outskirts of massive galaxies or in dwarf galaxies
where a significant discrepancy between the visible and dynamical masses is observed.
Thus, to catch a glimpse of new field(s) the targets should be selected 
either far from galactic centres or in dwarfs.

In the present work, we search for a new field-induced coupling in molecular clouds within
the Large and Small Magellanic Clouds (LMC and SMC).
Both of them are bright dwarfs detached from the solar system by $\sim 50$ kpc
and $\sim 60$ kpc, respectively. 
In the SMC, the DM component is believed to dominate the overall mass budget
at all distances from the centre, 
$\rho_{\rm b}/\rho_{\scriptscriptstyle{\rm DM}} < 0.5$ (Di Teodoro \etal\ 2019). 
In the LMC, the contribution of the DM halo is not well constrained 
because of complex kinematics of stars and gas 
which prevents the detailed rotation curve decomposition 
(Vasiliev 2018; van der Marel \& Kallivayalil 2014), 
but at galactocentric distances $R > 3$ kpc the input of DM definitely becomes significant, 
$\rho_{\rm b}/\rho_{\scriptscriptstyle{\rm DM}} < 1$ (Buckley \etal\ 2015). 

If new fields do exist and couple to the standard matter, 
then the fundamental coupling constants, primarily the proton-to-electron mass
ratio $\mu = m_{\rm p}/m_{\rm e}$ and possibly the
fine structure constant $\alpha = e^2/\hbar c$, 
are predicted to vary (for a review, see, e.g., Safronova \etal\ 2018; 
Kozlov \etal\ 2018; Brax 2014; Uzan 2011).
There are many atomic and molecular transitions which are highly sensitive to small changes in $\mu$ and 
$\alpha$ or in their combinations (Kozlov \& Levshakov 2013), 
i.e., relative offsets in the transition frequencies
induced by alleged changes in $\mu$ and/or $\alpha$ are large enough to be measured 
by spectroscopic methods which provide an unprecedented precision. 

As yet, spectroscopy~-- both at optical and radio bands~-- was used 
to probe the time and space dependence of $\mu$ and $\alpha$ or their combinations 
both for cosmological and local objects 
(e.g., Ubachs 2018; Kanekar \etal\ 2018; Gupta \etal\ 2018; Levshakov \etal\ 2017). 
Up to now, the most stringent limits on fractional change in $\mu$, 
$\Delta\mu/\mu = (\mu_{\rm obs} - \mu_{\rm lab})/\mu_{\rm lab}$,
and in $\alpha$,
$\Delta\alpha/\alpha = (\alpha_{\rm obs} - \alpha_{\rm lab})/\alpha_{\rm lab}$,
are as follows:  
$\Delta\mu/\mu = (-3.0\pm6.0)\times10^{-8}$ in one absorption-line system at $z = 0.89$ 
(Kanekar \etal\ 2015; Marshall \etal\ 2017);
$\Delta\mu/\mu = (-3.5\pm1.2)\times10^{-7}$ in one absorber at $z = 0.69$ (Kanekar 2011);
$\Delta\mu/\mu = (-3.3\pm1.9)\times10^{-8}$ in the dark cloud core L1498 in the MW disk (Dapr\`a \etal\ 2017);
$\Delta\mu/\mu = (-2.0\pm1.0)\times10^{-8}$ averaged over dark cloud cores 
in the MW disk within 300 pc from the Sun (Levshakov \etal\ 2010a,c, 2013).
As for $\alpha$-variations, they were constrained at a much less sensitive limit:
$\Delta\alpha/\alpha = (0.1\pm1.7)\times10^{-6}$ (Quast \etal\ 2004)
and
$\Delta\alpha/\alpha = (-0.1\pm0.8)\times10^{-6}$ (Levshakov \etal\ 2006) at $z = 1.15$;
$\Delta\alpha/\alpha = (-1.5\pm2.6)\times10^{-6}$ at $z = 1.58$ (Agafonova \etal\ 2011);
$\Delta\alpha/\alpha = (1.3\pm2.4)\times10^{-6}$ at $z = 1.69$ (Molaro \etal\ 2013); 
$\Delta\alpha/\alpha = (-1.4\pm0.9)\times10^{-6}$ at $z = 1.15$ (Kotu$\check{\rm s}$ \etal\ 2017);
$\Delta\alpha/\alpha = (3.3\pm2.9)\times10^{-6}$ at $z = 1.84$ (Bainbridge \& Webb 2017).
All estimates are given at $1\sigma$ statistical significance.

These numbers show that competitive estimates of $\Delta\mu/\mu$ and
$\Delta \alpha/\alpha$ should be at the level of a few $10^{-7}$
which requires observations with a high spectral resolution and a high signal-to-noise (S/N) ratio. 
In this sense the best currently available data are the observations of the molecular clouds in the LMC 
and the SMC performed by the {\it Herschel Space Observatory}\footnote{{\it Herschel} 
is an ESA space observatory with science instruments
provided by European-led Principal Investigator consortia and with
important participation from NASA.}. 

As trial transitions we chose the rotational emission lines of CO, and the fine structure (FS) emission
lines of [C\,{\sc i}] and [C\,{\sc ii}]. 
The use of a combination of molecular rotational and atomic fine structure transitions
to test the variability of the fundamental physical constants
was first suggested by Levshakov \etal\ (2008).
This approach being applied to constrain $\mu$- and/or $\alpha$-variations in Galactic
and extragalactic objects has proved itself as
a powerful tool for probing the stability of the fundamental constants
with an accuracy widely exceeded that of optical spectral measurements
(Levshakov \etal\ 2010b, 2017). 
Another advantage of the [C\,{\sc i}], [C\,{\sc ii}], and CO lines is that they are 
the most abundant species observed both in local and high redshift molecular systems
and, hence, are convenient to make primary estimations and to select perspective targets.

\section{Method}
\label{Sec2}

The frequencies of the rotational lines of light molecules are independent of $\alpha$, but sensitive 
to $\mu$, whereas the fine structure transition frequencies are 
proportional to $\alpha^2$ and independent of $\mu$.
Thus, comparing the observed frequencies of rotational and fine structure lines
we estimate the value of $\Delta F/F$
with $F = \mu\alpha^2$ and $\mu = m_{\rm p}/m_{\rm e}$
(Levshakov \etal\ 2008, 2010b).
Converting the frequency scale, $\nu$, into the velocity scale, $V$, by 
\begin{equation}
V/c = 1 - \nu_{\rm obs}/\nu_{\rm lab} ,
\label{Eq1a}
\end{equation}
at $V \ll c$, the fractional change $\Delta F/F$ is calculated as:
\begin{equation}
{\Delta F}/{F}  = {\Delta \mu}/{\mu} + 2{\Delta \alpha}/{\alpha} = {\Delta V}/{c},
\label{Eq1}
\end{equation}
where $\Delta V = V_{\rm rot} - V_{\rm fs}$ is the difference between the radial velocities of the CO
rotational line and the [C\,{\sc i}] or [C\,{\sc ii}] FS line, 
and the radial velocity is the velocity of the line centre related to the observed
and laboratory frequencies, $\nu_{\rm obs}$ and $\nu_{\rm lab}$.

The line centre is determined through the fitting of a set of Gaussian components to the observed line.
If the line is symmetric (one component profile), the line centre is simply a fitting parameter
and its statistical error $\sigma_{\rm stat}$ is calculated by the standard procedure employing 
the covariance matrix.

If, however, the line is asymmetric (multicomponent profile), then its centre is defined as a point where
the first order derivative of the fitting curve is equal to zero.
Employing a common approach (e.g., Savitzky \& Golay 1964), we draw a parabola
through three points $\{x_1,y_1; x_2,y_2; x_3,y_3\}$ of the fitting curve $y(x)$
which include the intensity peak, and calculate
the line centre $x_0$ and its error $\sigma_{\rm stat}$ as
\begin{equation}
x_0 = \frac{x_1+x_2}{2} - \frac{(y_2-y_1)\Delta_{\rm ch}}{y_1-2y_2+y_3} ,
\label{Eq10}
\end{equation}
and
\begin{eqnarray*}
\lefteqn{ \sigma_{\rm stat} = } \\
& &{} \frac{\sigma_{\rm rms}\cdot \Delta_{\rm ch}}{(y_1 - 2y_2 + y_3)^2}
\sqrt{(y_3-y_2)^2 + (y_1-y_3)^2 + (y_2-y_1)^2}, 
\end{eqnarray*}
where $\sigma_{\rm rms}$ is the average noise of a given spectrum, 
and the channel width $\Delta_{\rm ch} = x_2-x_1 = x_3-x_2$.

Under real conditions the spatial distributions of different species
do not trace each other exactly. This leads to additional velocity shifts
between the line centres due to different kinematics of the emitting 
regions~-- the so-called Doppler noise (DN). 
In a single measurement, the DN may either mimic or obliterate a sought-for signal in $\Delta F/F$.
However, the DN is supposed to be random and normally distributed, i.e., it has 
a zero mean and a finite variance, and, thus,
it can be significantly reduced by averaging over a set of measurements.

\section{Data}
\label{Sec3}

We use deep observations of fine structure transitions of \CI\
$^3$P$_2$--$^3$P$_1$ (herein [C\,{\sc i}]), 
\CII\ $^2$P$_{3/2}$--$^2$P$_{1/2}$ and the
$J$=7--6 rotational transition of CO towards LMC and SMC.  
These observations were performed with the
high resolution Heterodyne Instrument in the Far Infrared (HIFI)
(de Graauw \etal\ 2010) on board the {\it Herschel Space Observatory} (Pilbratt \etal\ 2010).
This data is complemented with observations of the $J$=1--0 and $J$=3--2 transitions
of $^{12}$CO and $^{13}$CO from the ATNF Mopra\footnote{The Mopra radio
telescope is part of the Australia Telescope which is funded by the
Commonwealth of Australia for operation as a National Facility managed by CSIRO.} 
and APEX\footnote{This publication is based in part on data
acquired with the Atacama Pathfinder Experiment (APEX). APEX is a 
collaboration between the Max-Planck-Institut f\"ur Radioastronomie, the
European Southern Observatory, and the Onsala Space Observatory.}
telescopes respectively. 
We have retrieved the final reduced {\em User Provided Data Product} from the
Herschel archive for the purpose of our analysis.  
The full description of observational data used in the present work is given by
Pineda \etal\ (2017).

The baseline for each spectrum
was defined by selecting spectral windows without emission lines and/or
noise spikes and then calculating the mean main beam brightness
temperature $T_{\rm mb}$ along with its rms uncertainty $\sigma$ for
each spectral window. 
Using spline interpolation through this set of pairs
\{$T_i, \sigma_i$\} we obtained a baseline which was subtracted from
the spectrum.
Since the rms uncertainties $\sigma_i$ are approximately
identical for all windows, their mean value $\sigma_{\rm rms}$ was assigned to
the whole spectrum.

\begin{figure*}
 \includegraphics[width=14.0cm]{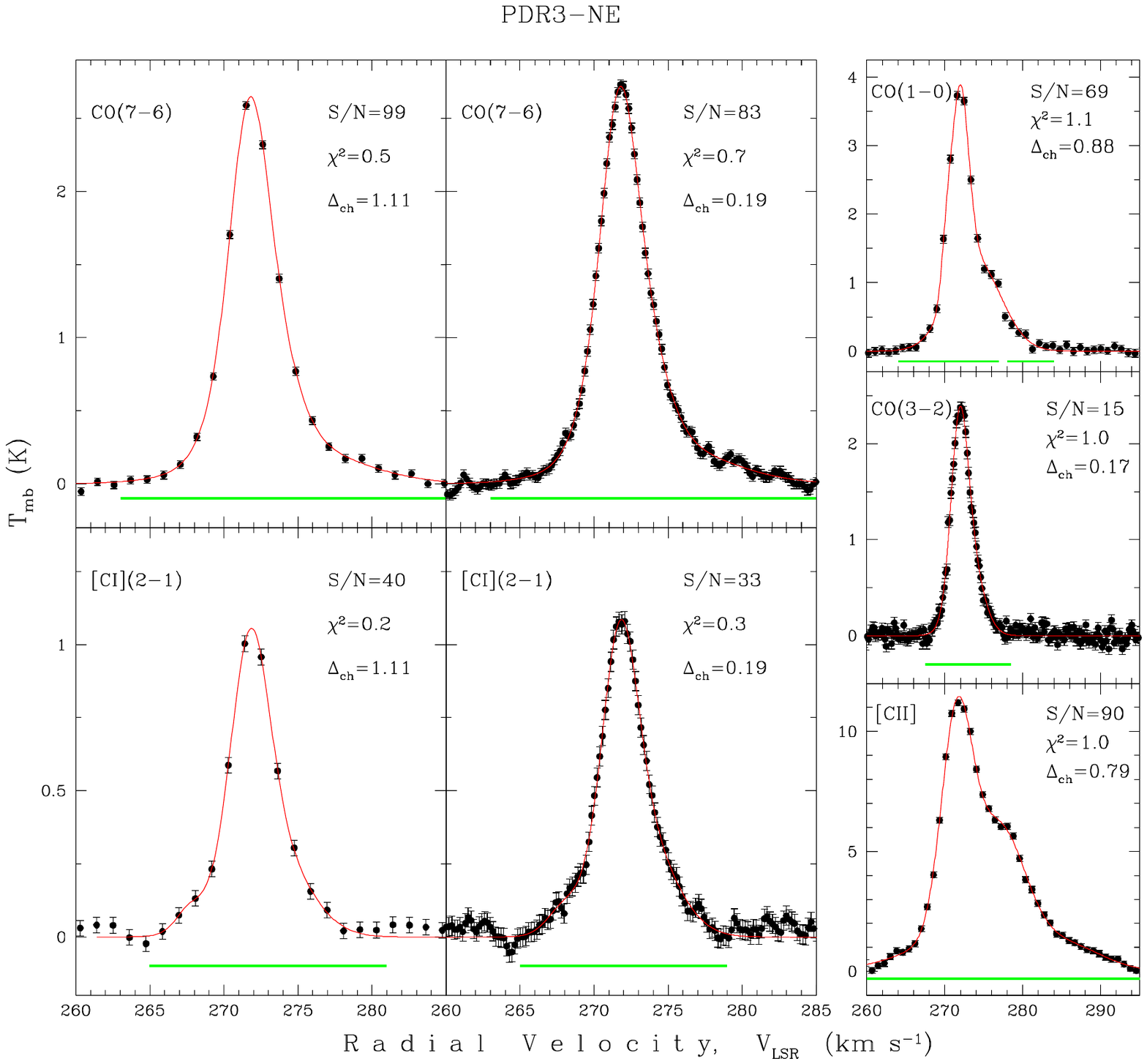}
\vspace{-3.0cm}
\caption{
Observed spectra (dots with $1\sigma$ error bars) towards the molecular cloud PDR3-NE
in the LMC (see Table~\ref{T1}). 
Radial velocities are given in \kms\ relative to the Local Standard of Rest (LSR)
and the line intensities are in units of the telescope mean beam temperature $T_{\rm mb}$ (K).
For each spectrum 
the signal-to-noise ratio (S/N) per channel at the maximum intensity peak, 
$\chi^2$ per degree of freedom, and the channel width $\Delta_{\rm ch}$ (in \kms) are indicated.
The fitting curves are shown by red.
The horizontal green lines mark spectral ranges included in the fitting procedure.
}
 \label{Fg1}
\end{figure*}

\begin{figure*}
 \includegraphics[width=14.0cm]{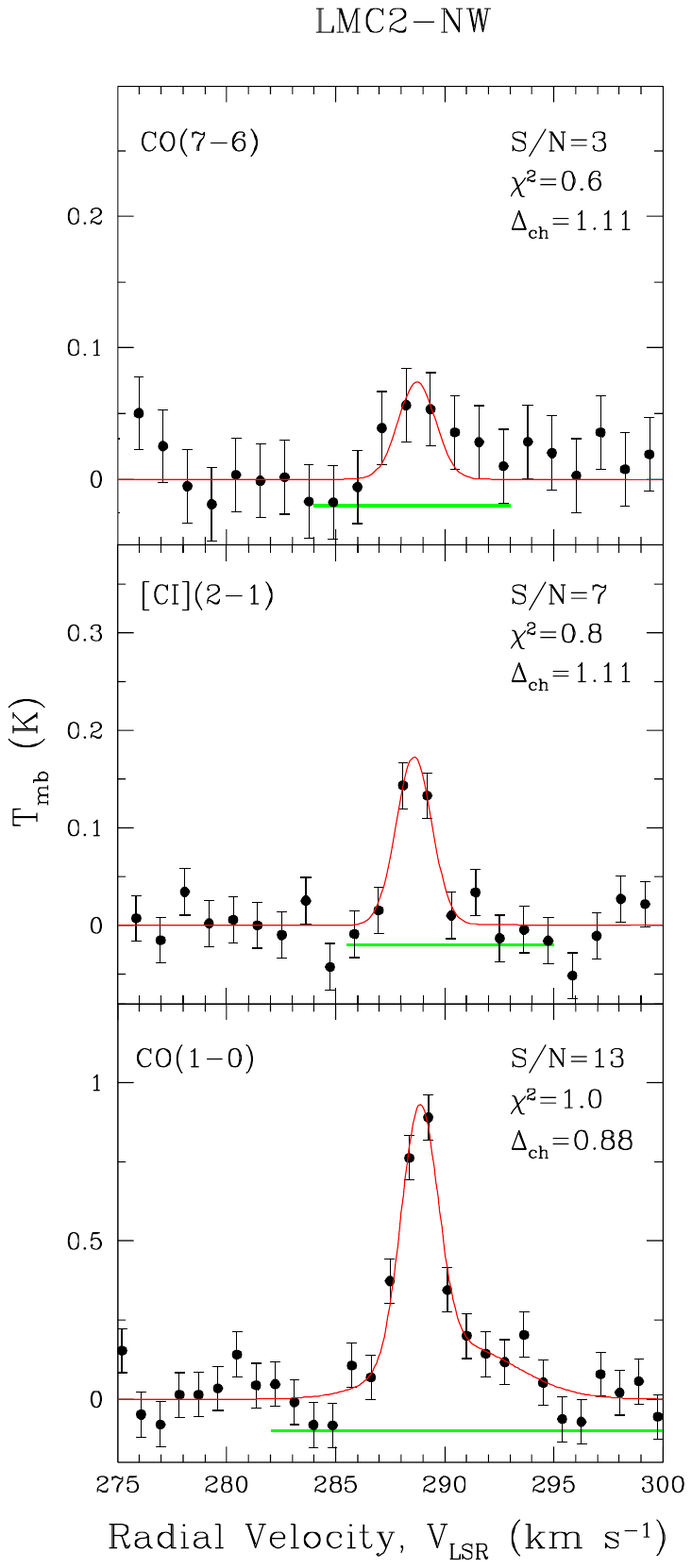}
\vspace{-3.0cm}
 \caption{Same as Fig.~\ref{Fg1}, but for the molecular cloud LMC2-NW in the LMC. }
 \label{Fg3}
\end{figure*}

\begin{figure*}
 \includegraphics[width=14.0cm]{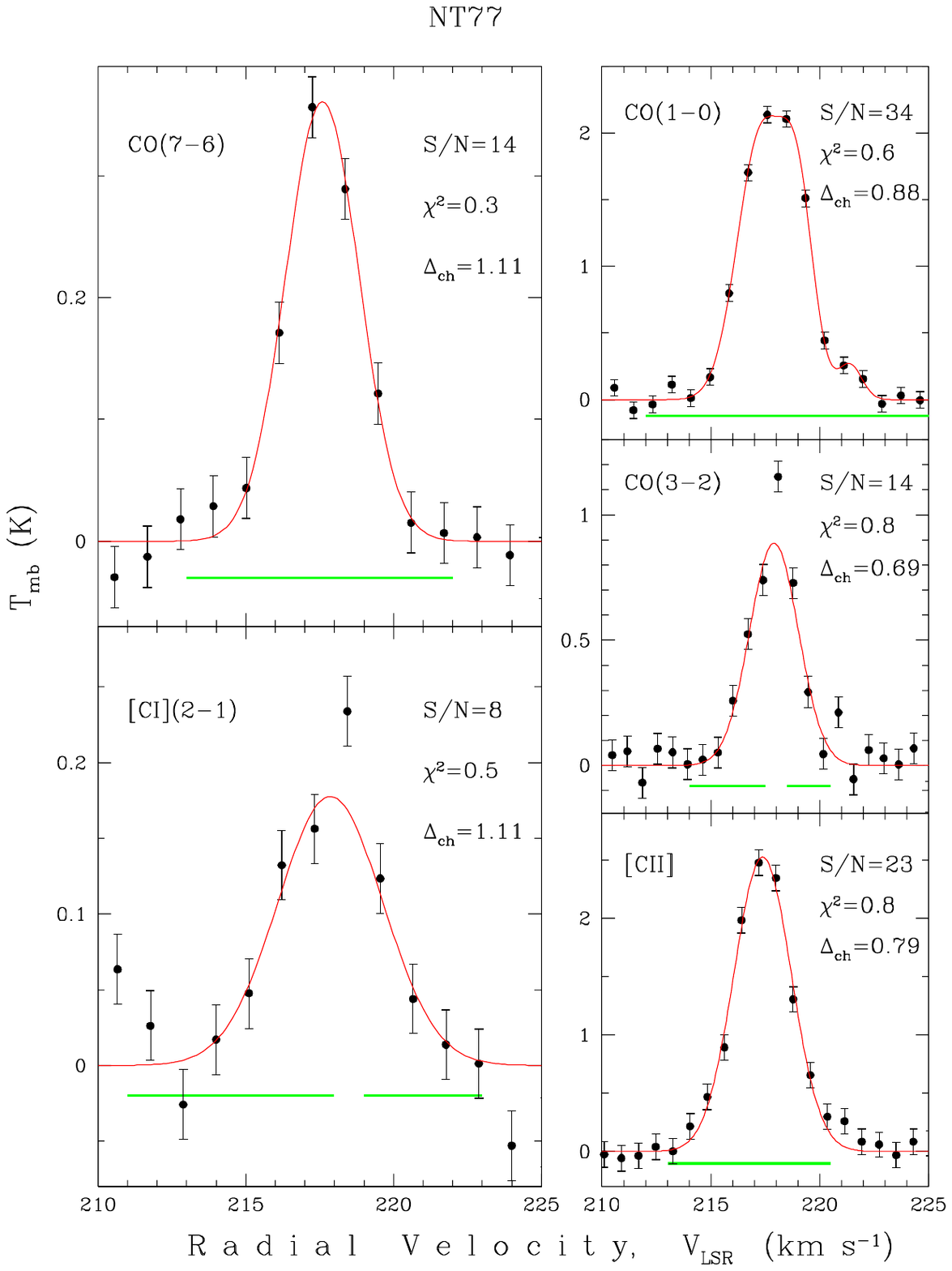}
\vspace{-3.0cm}
 \caption{Same as Fig.~\ref{Fg1}, but for the molecular clouds NT77 in the LMC. }
 \label{Fg4}
\end{figure*}

\begin{figure*}
 \includegraphics[width=14.0cm]{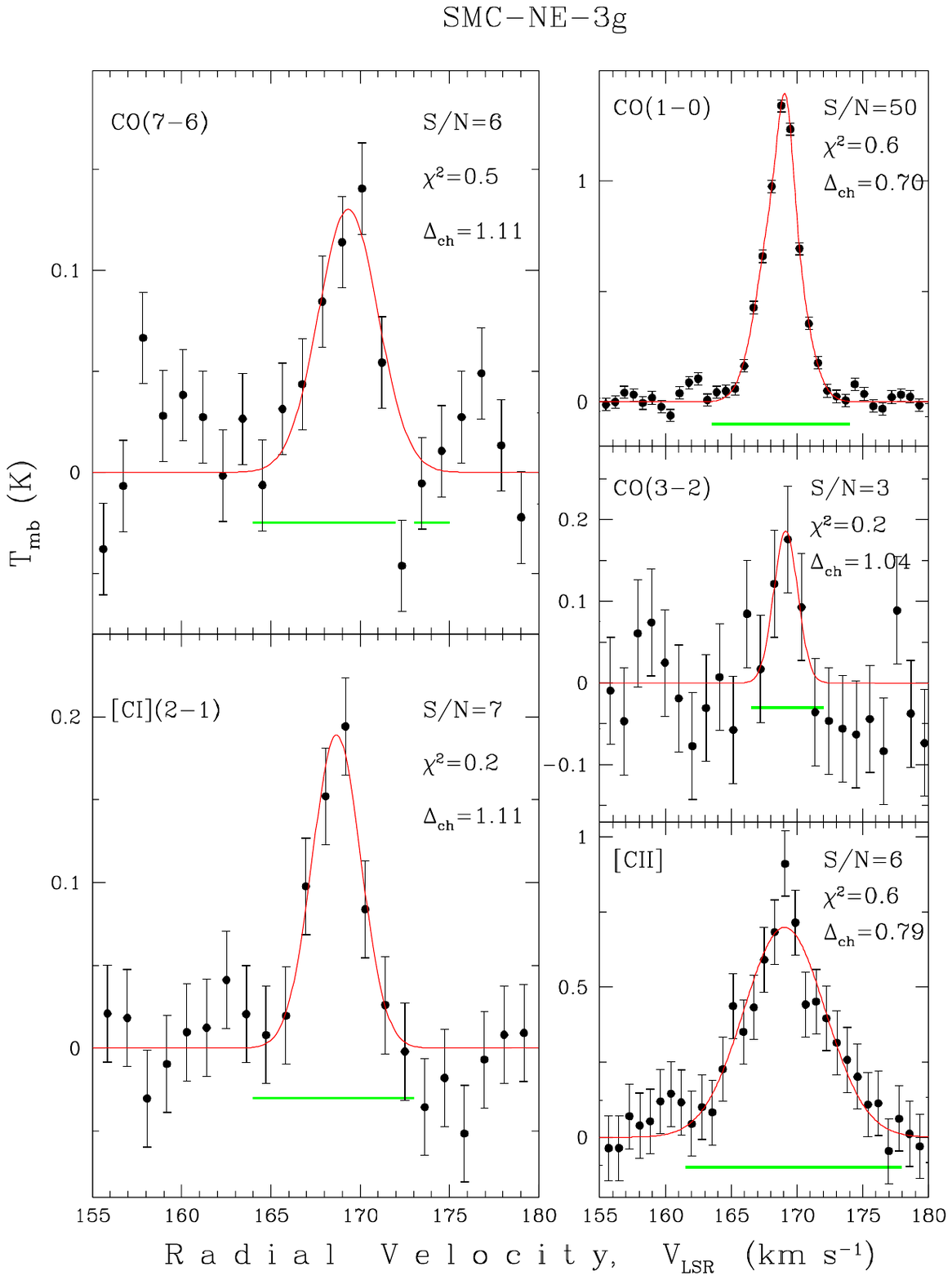}
\vspace{-3.0cm}
 \caption{Same as Fig.~\ref{Fg1}, but for the molecular cloud SMC-NE-3g in the SMC.}
 \label{Fg5}
\end{figure*}

\begin{figure*}
 \includegraphics[width=14.0cm]{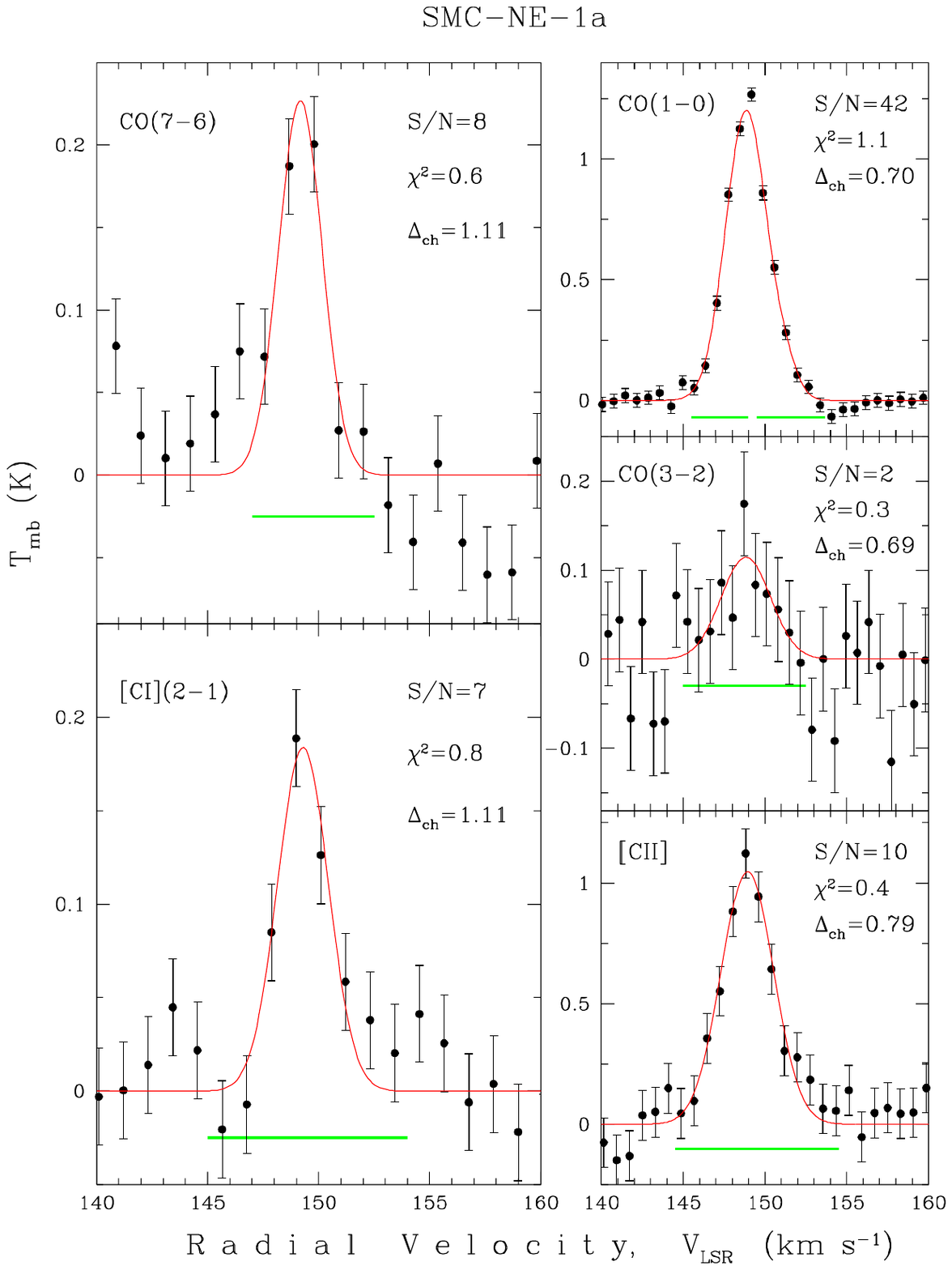}
\vspace{-3.0cm}
 \caption{Same as Fig.~\ref{Fg1}, but for the molecular cloud SMC-NE-1a the SMC.}
 \label{Fg6}
\end{figure*}

\begin{figure*}
 \includegraphics[width=14.0cm]{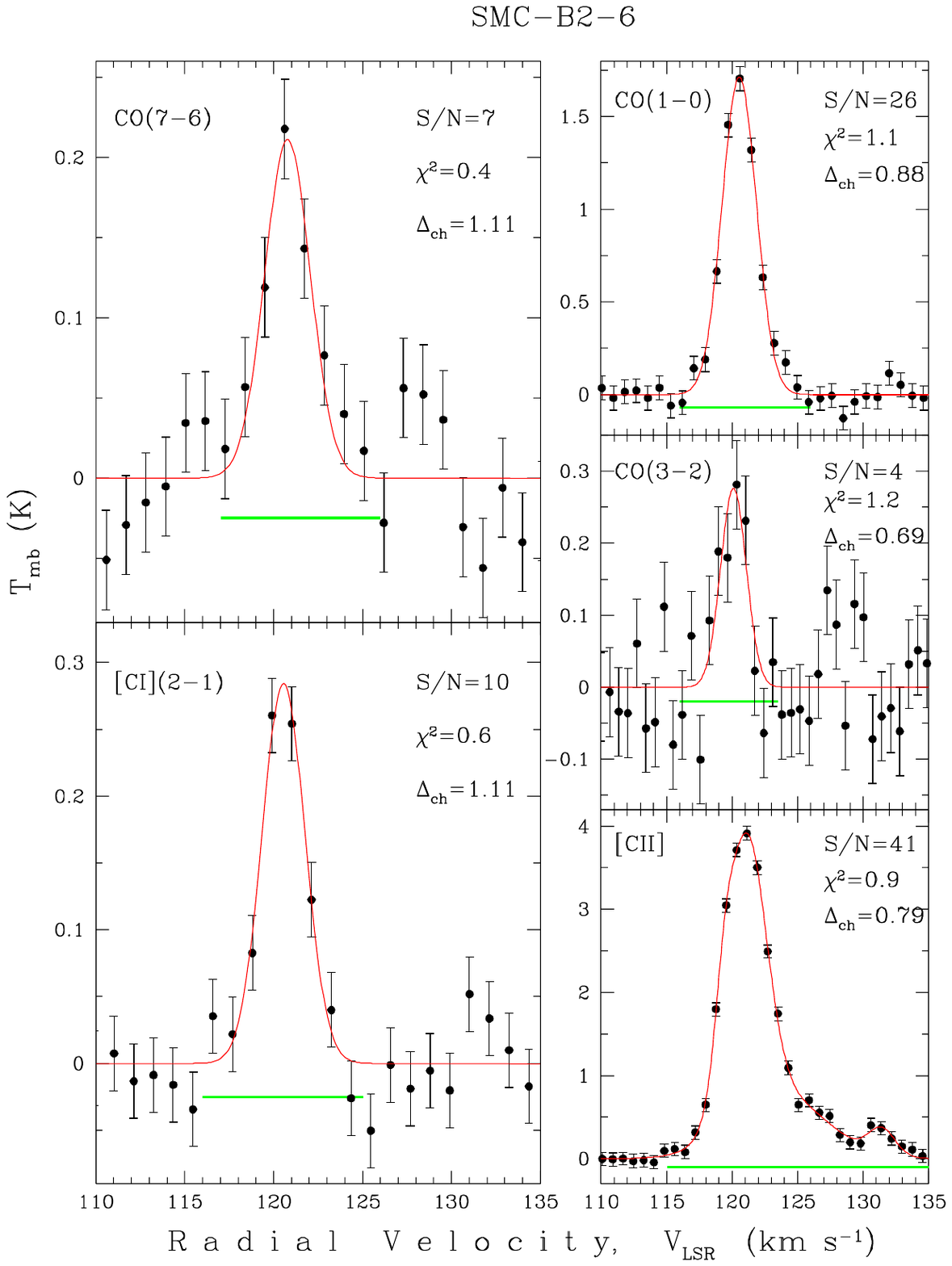}
\vspace{-3.0cm}
 \caption{Same as Fig.~\ref{Fg1}, but for the molecular cloud SMC-B2-6 in the SMC.}
 \label{Fg7}
\end{figure*}

\begin{figure*}
 \includegraphics[width=14.0cm]{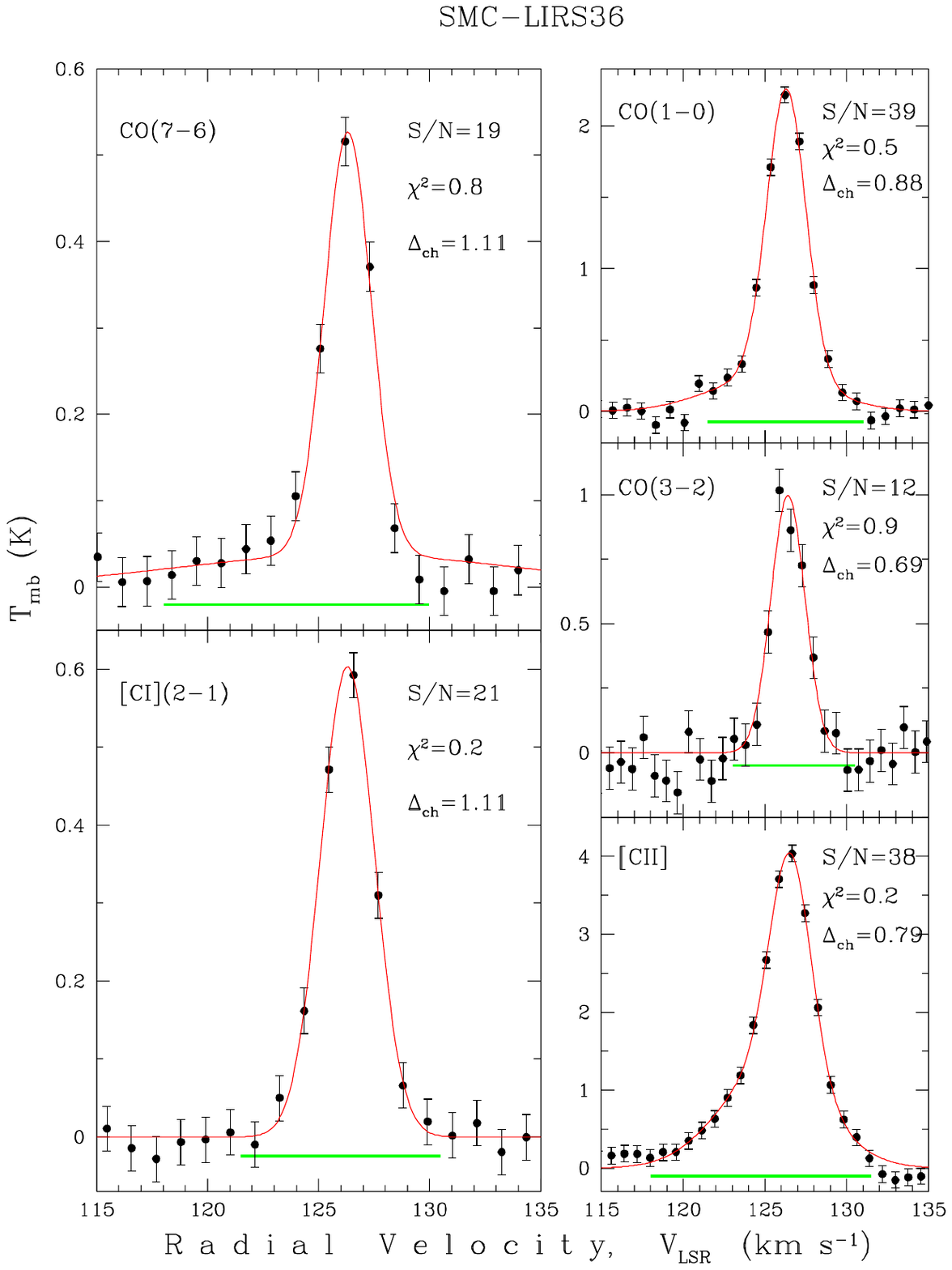}
\vspace{-3.0cm}
 \caption{Same as Fig.~\ref{Fg1}, but for the molecular cloud SMC-LIRS36 in the SMC.}
 \label{Fg8}
\end{figure*}

\begin{figure*}
 \includegraphics[width=14.0cm]{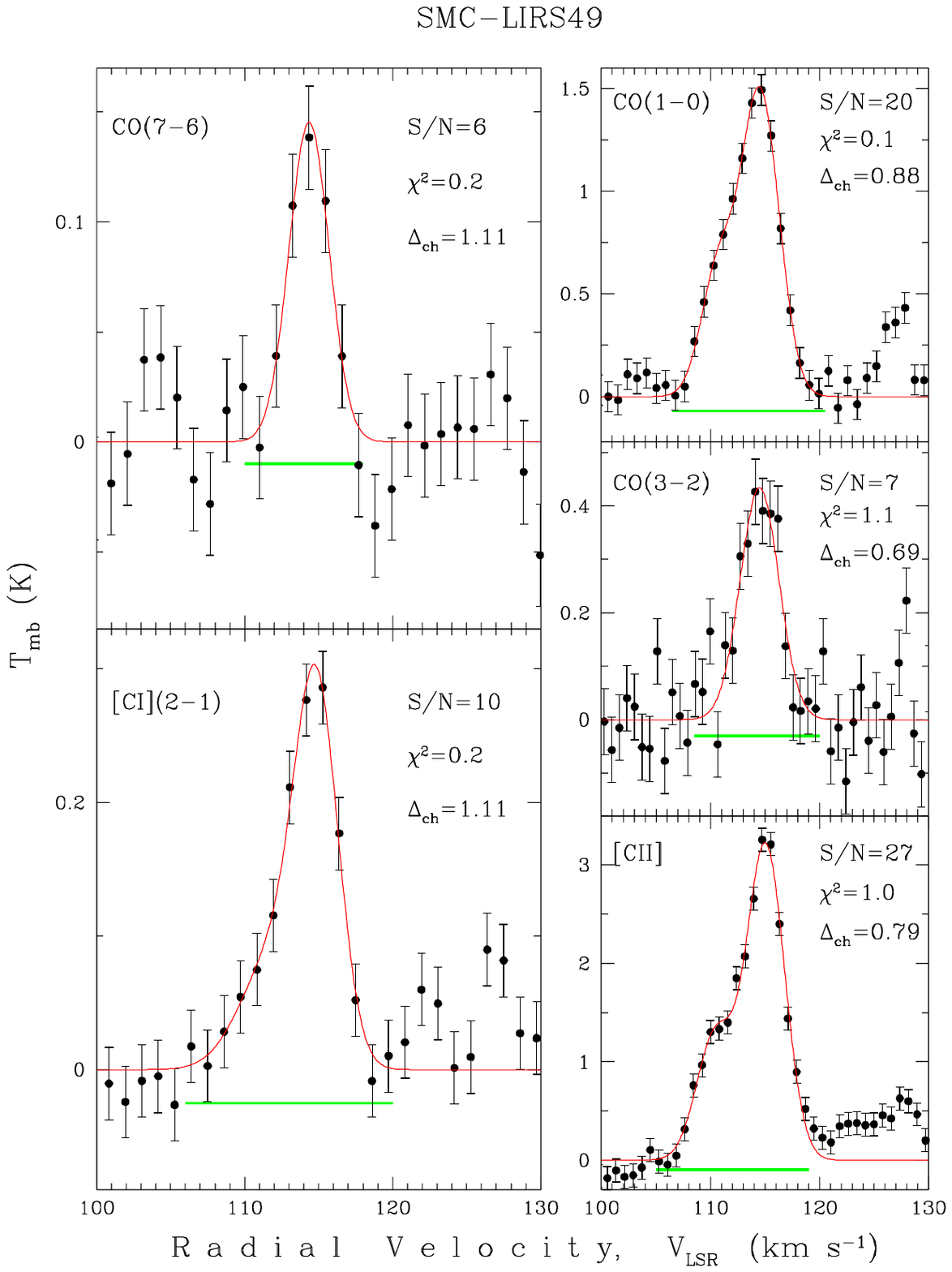}
\vspace{-3.0cm}
 \caption{Same as Fig.~\ref{Fg1}, but for the molecular cloud SMC-LIRS49 in the SMC.}
 \label{Fg9}
\end{figure*}

Table~\ref{T1} summarizes for all spectral lines
considered here the laboratory frequencies and their uncertainties
taken from the papers cited in the last column,
the corresponding systematic errors in velocity ($\sigma_{\rm sys}$), and the
beamsizes of the observations. 
The spectral lines are analyzed at different spectral resolutions
which are given in Table~\ref{T2} and indicated in Figs.~\ref{Fg1}-\ref{Fg9}.
The errors $\sigma_{\rm sys}$ determine the limiting
precision with which the line centres can be measured. 
In particular,
$\sigma$($\Delta$F/F)$_{lim}$ = 2$\times$10$^{-8}$ for the pair \CI/CO(7--6).

\section{Results and discussion}
\label{Sec4}

\begin{figure*}
 \includegraphics[width=14.0cm]{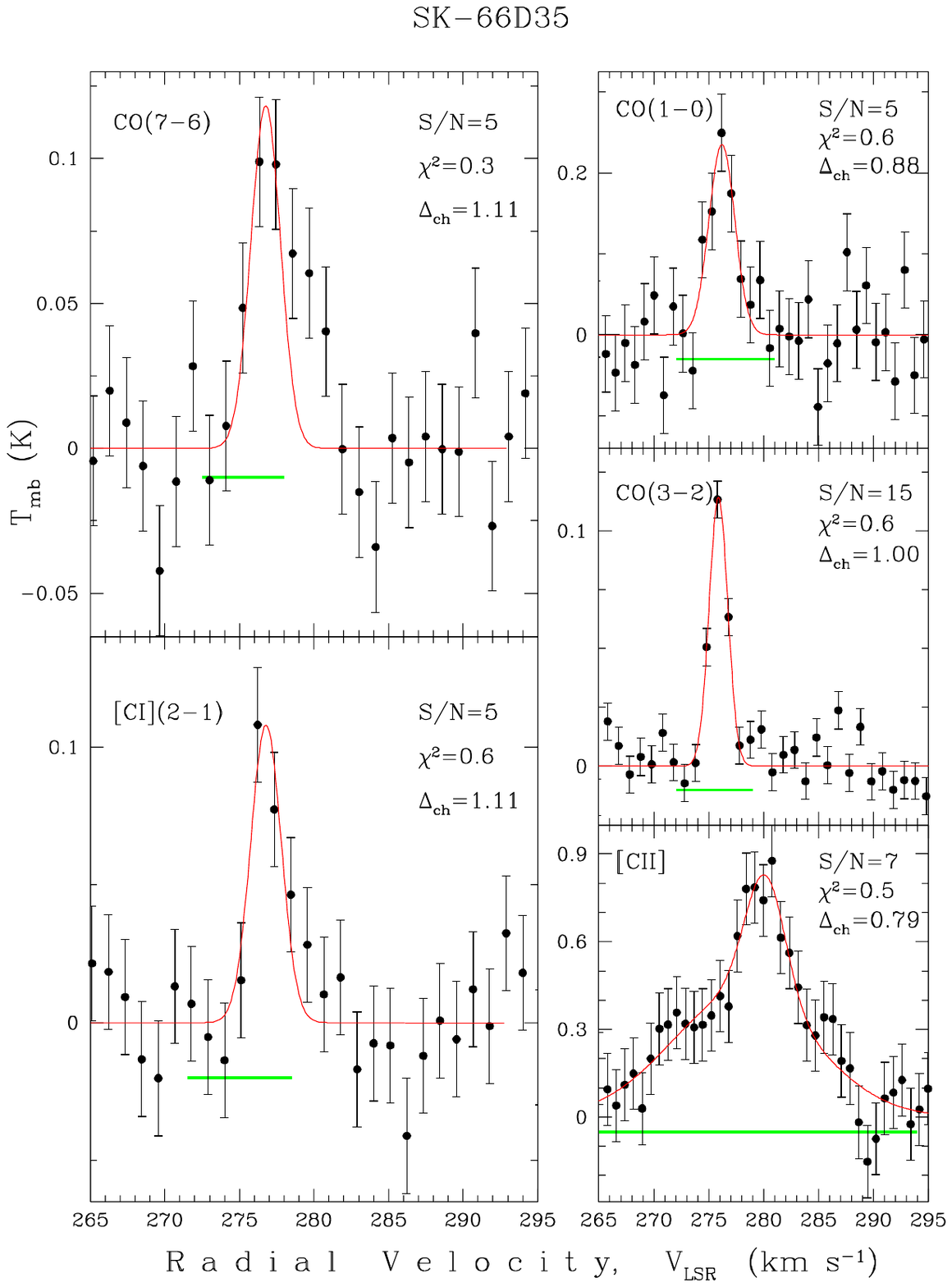}
\vspace{-3.0cm}
 \caption{Same as Fig.~\ref{Fg1}, but for the molecular cloud SK-66D35 in the LMC.}
 \label{Fg2}
\end{figure*}

The measured lines centres, $V_{\scriptscriptstyle\rm LSR}$, and
their $1\sigma$ statistical errors are listed in Table~\ref{T2}.
Figures~\ref{Fg1}-\ref{Fg9} show the observed spectra (dots with $1\sigma$ error bars)
and model curves (red solid lines) obtained by the fitting procedure.
 
In order to analyze a homogeneous sample we combine the 1.11 \kms\ channel width 
data (index ``$b$'' in Table~\ref{T2}), while the PDR3-NE 0.19 \kms\ channel width data
(index ``$c$'' in Table~\ref{T2}) are considered separately.
Since the sample size is small, $n = 9$, the Student's $t$-test should be utilized
to calculate the sample variance.
It is also to note that
the errors of the velocity offsets given in parenthesis in Col.4, Table~\ref{T2}, range
between 0.08 \kms\ and 0.7 \kms, whereas the velocity offset corresponding to 
$\Delta F/F \sim 10^{-7}$~-- the value set by the available limits on 
\dmm\ cited in Sect.~\ref{Sec1}~-- should be $\sim 0.03$ \kms. 
Thus, the data are clearly noise-dominated. 
Then the errors of $\Delta V$ cannot be used as weights in calculations
of the sample mean and its variance because they are the errors of the noise amplitude and
not the errors of the anticipated signal.
That is why all statistical estimates are made with simple (unweighted) averaging
of data points.

The velocity offset
between lines of different elements is the sum of two components:
\begin{equation}
\Delta V = \Delta V_{\scriptscriptstyle F} + \Delta V_{\scriptscriptstyle D}\ ,
\label{Eq4}
\end{equation}
where $\Delta V_{\scriptscriptstyle F}$ is
a regular velocity shift due to variation in $F$, and
$\Delta V_{\scriptscriptstyle D}$ is a random component caused by DN.
Supposing a normally distributed DN and averaging over the total sample
we obtain the mean value  
\begin{equation}
\langle \Delta V \rangle  =  \langle \Delta V_{\scriptscriptstyle F} \rangle , 
\label{Eq5}
\end{equation}
and the error of the mean
\begin{equation}
\sigma_{\scriptscriptstyle \langle \Delta V \rangle}  = 
t_{p,\nu} \sigma_{\scriptscriptstyle \Delta V}/\sqrt{n} ,
\label{Eq5a}
\end{equation}
where $n$ is the sample size, and $t_{p,\nu}$ is the Student's $t$-test coefficient for 
the probability $p = 0.7$ ($1\sigma$) and the number of degrees of freedom $\nu = 8$.

The spectral lines CO(7--6) and \CI\ are of {\em prime interest} 
to us since these transitions have nearby frequencies and were observed
in the same band, i.e., simultaneously and with identical angular resolution
(see Table~\ref{T1}). This eliminates part of the DN which can be caused by different beam
filling factors for both emitting regions\footnote{At the LMC and SMC
distances $\theta$ = 1\arcsec\ correspond to $R\simeq$ 0.2--0.3\,pc.}.
We note that different spatial morphologies and/or kinematic structure
of the tracers within a single given beam size give rise to another
part of DN which is of course not removed by such choice of lines.

The mean, $\langle \Delta V \rangle$,
and the standard deviation, $\sigma_{\scriptscriptstyle \Delta V }$,
of the velocity offsets between the $n = 9$ pairs of
CO(7--6) and [C\,{\sc i}] lines from the 1.11 \kms\ channel width data
are $\langle \Delta V \rangle = -0.02$ \kms\ and
$\sigma_{\scriptscriptstyle \Delta V } = 0.2$ \kms. 
The non-parametric estimates for the centre (median) and the dispersion
($=1.48MAD$, Median Absolute Deviation) give $\Delta V_{\rm med} = -0.01$ \kms\
and $\sigma_{{\scriptscriptstyle {\Delta V}}_{\rm med}} = 0.3$ \kms. These
estimates are robust against large deviations from the sample mean and are usually
used to detect outliers. Since here $\sigma_{{\scriptscriptstyle {\Delta V}}_{\rm med}} >
\sigma_{\scriptscriptstyle \Delta V }$, the sample can be considered
as homogeneous.

For comparison, averaging over the sample of offsets
between the same fine structure line [C\,{\sc i}]  
from the {\it Herschel} data and rotational lines CO(1--0) and CO(3--2) 
observed with other facilities and with lower/higher angular resolutions, respectively, we obtain
for the [C\,{\sc i}]/CO(1--0) pair
$\langle \Delta V \rangle = -0.09\pm0.3$ \kms\ and for  [C\,{\sc i}]/CO(3--2) 
$\langle \Delta V \rangle = -0.003\pm0.6$ \kms,
i.e., as expected, these estimates are noticeably more dispersed.
Taking this into account, all conclusions below refer to the measurements 
on the base of [C\,{\sc i}]/CO(7--6). 

The use of the [C\,{\sc ii}] line is less favorable than of [C\,{\sc i}] 
since [C\,{\sc ii}] typically shows wider line profiles or additional velocity components
whereas the [C\,{\sc i}] and CO profiles are usually similar (Okada \etal\ 2019).
This agrees with the result of Pineda \etal\ (2017) that
C$^+$ is the dominant gas-phase form of carbon associated with 
photodissociation regions (PDRs)~-- neutral regions where chemistry
and heating are regulated by the far-UV photons (Hollenbach \& Tielens 1999).
Photons with energy $E > 11.1$ eV dissociate CO into atomic carbon and oxygen in PDRs.
Since the C$^0$ ionization potential of 11.3 eV is quite close to the CO dissociation
energy, neutral carbon can be quickly ionized. This suggests the
chemical stratification of the PDR in a row C$^+$/C$^0$/CO with
increasing depth from the surface of the PDR.
That is why C$^0$ and CO represent only a small fraction of PDRs, namely 
that where neutral gas is well shielded from the far-UV photons.

For instance, in our dataset wider and complex [C\,{\sc ii}] profiles with additional velocity
components are observed practically in all systems, 
with SK-66D35 (Fig.~\ref{Fg2}) being the most evident case
where the peak of the [C\,{\sc ii}] emission is shifted by $\sim 3-4$ \kms\
with respect to the other lines (see Table~\ref{T2})\footnote{Pineda \etal\ (2017)
consider the [C\,{\sc ii}] and [C\,{\sc i}]/CO emission as arising from two different sources 
SK-66D35-1 and SK-66D35-2.}. 
In the present work we use the [C\,{\sc ii}] as well as low-$J$ CO lines only to 
constrain the kinematic structure within the molecular clouds. 

Now return to our sample of velocity offsets $\Delta V$ between [C\,{\sc i}] and CO(7--6) lines.
In spite of its smallness, the sample is consistent with a normal distribution on formal criteria 
(sample values of the mean absolute deviation, asymmetry and kurtosis are well within $1\sigma$
limits for a normal distribution).
This allows us to calculate the mean and the error of the mean as
given in (\ref{Eq5}, \ref{Eq5a}): $\langle \Delta V \rangle = -0.02\pm0.07$ \kms. 
The systematic error due to uncertainties in the laboratory
frequencies is almost an order of magnitude lower, so that the value of 0.07 \kms\ 
can be considered as the total measurement error\footnote{For comparison, 
the weighted mean and the standard deviation for the same sample  
(weights calculated as inverse squares of errors) are
$\langle \Delta V \rangle = -0.05$ \kms\ and
$\sigma_{{\scriptscriptstyle \Delta V}} = 0.12$ \kms.
This demonstrates that the weighted standard deviation is underestimated.}.
Being expressed in terms of $\Delta F/F$, this error restricts the variability
of $F$ at the level  $|\Delta F/F| < 2\times10^{-7}$.

For one target in the LMC~-- PDR3-NE~--
apart from measurements with a channel size of 
$\Delta_{\rm ch} = 1.11$ \kms\ common for all targets,
there are also measurements of the [C\,{\sc i}] and CO(7--6) lines with 
$\Delta_{\rm ch} = 0.19$ \kms\ (see Table~\ref{T2}). 
Such a high spectral resolution makes it possible
to calculate the line centres with a statistical error of only $0.01-0.02$ \kms. 
For the velocity offset we get in this case $\Delta V = -0.05\pm0.02$ \kms\
and, correspondingly,
$|\Delta F/F| = (-1.7\pm0.7)\times10^{-7}$.
Although consistent with the overall $\Delta V$ estimate, here the negative
offset becomes statistically significant.
This target is located at the projected distance of about 4~kpc
from the LMC centre, i.e., it falls in the region where DM dominates in the mass budget. 
Unfortunately, with the present data we cannot deduce whether this offset is entirely (or partly) caused
by variations in $F$ or indeed by different velocity fields in the CO and [C\,{\sc i}] 
emitting regions. 
Namely, the [C\,{\sc i}] and [C\,{\sc ii}] line centres
taken with the {\it Herschel Space Observatory} 
coincide within the $1\sigma$ uncertainty interval in spite of the 
differences in the angular resolutions and the line profile shapes.
This can be interpreted as a cloud with a very compact and homogeneous core~--
as already was discussed above, generally  [C\,{\sc ii}] emitting regions are much larger in size than 
those of [C\,{\sc i}] and, hence, their velocity centroids may diverge.
On the other hand,
the centres of the rotational CO lines observed with other facilities show a large scatter 
(e.g., the offset $\Delta V \simeq 1.1$ \kms\ between CO(3--2) and CO(7--6) 
observed with similar resolution, $\Delta_{\rm ch} \simeq 0.2$ \kms, Table~\ref{T2}) 
which hardly can be explained. 
Clearly, the kinematic structure of this perspective target should be studied in more detail.

Assuming that the variations in $\mu$ may exceed those of $\alpha$
(see, e.g., Langacker \etal\ 2002; Flambaum 2007; Uzan 2011; Brax 2014),
we can transfer the obtained limits on $\Delta F/F$ to $\Delta \mu/\mu$.
A sign of \dmm\ is expected to be negative
since the additional coupling of the hypothetical non-standard field(s) to baryonic matter
may increase the mass of the electron $m_{\rm e}$, whereas
the proton mass $m_{\rm p}$, being determined mostly by the binding energy of quarks, 
remains practically unchanged.

It is interesting to compare the values of $\Delta \mu/\mu$ measured in the Magellanic Clouds where the
overall mass budget is dominated by DM with values from objects with different ratios of
baryonic matter to DM.
As was already mentioned above, baryons are supposed to constitute about 90\% 
of the dynamical mass at the position of the Sun.
For several nearby molecular cores (detached by $\la 300$ pc) the estimates of \dmm\ were obtained
in the framework of a program to test the chameleon screening. Strange enough, but in all measurements
a negative \dmm\ of a few $10^{-8}$ was reproduced in spite of using a variety of
molecular transitions and observing with different radio telescopes 
(Levshakov \etal\ 2010a,c; Dapr\`a \etal\ 2017). 
However, $10^{-8}$ is just the level of the systematic error which in the present case comprises 
both the limiting accuracy of the observing facilities and
the uncertainty in the laboratory frequencies of the employed molecular transitions 
(Levshakov \etal\ 2013). 
Thus, these results allow us to set only an
upper limit on possible changes in $\mu$ in media where baryonic matter dominates:  
$|\Delta \mu/\mu| < 3\times10^{-8}$.

There is one estimate of a comparable accuracy reported for an absorption
system at $z = 0.89$: \dmm\ = $(-3.0\pm6.0)\times10^{-8}$ (Kanekar \etal\ 2015; Marshall \etal\ 2017).
The estimate was obtained with different transitions of the same molecule CH$_3$OH, i.e.,
the influence of the DN and chemical segregation is minimal in this case.
The system probably originates in a spiral arm at a distance of $\sim 2$ kpc from the centre
of a massive spiral galaxy (Muller \etal\ 2006).
As expected, the limit on \dmm\ in the $z = 0.89$ system complies with that for the MW disk.

On the other hand,
Kanekar (2011) reports  $\Delta \mu/\mu = (-3.5\pm1.2)\times10^{-7}$ measured
on base of NH$_3$, CS and H$_2$CO molecular lines detected in a $z = 0.69$ absorber.
These absorption lines arise in the halo of a low luminosity dwarf galaxy and 
deep image observations do not reveal
any neighboring galaxy at the projected distance less than 65 kpc (Falomo \etal\ 2017), 
i.e., this system resembles the objects in the Magellanic Clouds.
However, the velocity offset between the above combination of molecules can contain a significant input
from DN~-- just as it is observed between the [C\,{\sc i}] and CO transitions used in our study.
To estimate and to eliminate the influence of DN, new measurements employing a complete palette of
suitable transitions are required.

\section{Summary and future prospects}
\label{Sec5}

We have used high resolution submillimeter spectra of the [C\,{\sc i}], [C\,{\sc ii}] and CO
emission lines from objects in the Magellanic Clouds to measure the velocity shifts
between molecular rotational lines and atomic fine structure lines. 
Such shifts can be (partly) attributed to the variations in the fundamental constants $\mu$ and $\alpha$ 
which in turn can be caused by coupling of the non-standard fields to ordinary matter.
The overall mass budget in the Magellanic Clouds is dominated by DM, thus making them
to be favourable targets to search for manifestations of such fields.

Using the {\it Herschel Space Observatory} observations of CO(7--6) 806.5 GHz and [C\,{\sc i}] 809.3 GHz 
transitions from nine molecular clouds in both the LMC and SMC and averaging 
over the whole sample we obtained 
$\langle \Delta V \rangle = -0.02\pm0.07$ \kms\ or, in terms of $\Delta F/F$,
$\Delta F/F = (-0.7\pm2.3)\times10^{-7}$, where $F = \mu \alpha^2$. 
However, for one object where the observations were carried out with a 5 times better
spectral resolution, the result was $\Delta V = -0.05\pm0.02$ \kms and, correspondingly,
$\Delta F/F = (-1.7\pm0.7)\times10^{-7}$.

Before attributing the measured velocity shifts
entirely (or partly) to changes in the fundamental constants
one has to exclude velocity components arising from the kinematic structure
of the observed clouds.

This can be done in different ways. One possibility is to study the gas velocity distribution
in greatest detail by involving other molecules.
For instance, CH is an intermediate molecule in gas-phase chemical reactions from C to CO
and, thus, its emission traces their common spatial distribution (Sakai \etal\ 2012).
The fundamental spin-rotational transitions of CH at
533 GHz and 537 GHz have already been considered
as probes for a possible variation of coupling constants (de Nijs \etal\ 2012).
Other CH transitions at lower frequencies 3.3 GHz and 0.7 GHz were discussed for the same tasks
in Kozlov (2009) and Truppe \etal\ (2013). 
Another important intermediate molecule for the production of CO is OH (Sakai \etal\ 2012)
which also can be used for mapping the velocity field in molecular clouds.

The second possibility is to use different transitions of the same molecule.
As is known, the methanol molecule has a complex microwave spectrum
with a large number of very strong lines which have different sensitivity to
$\mu$-variations (Jansen \etal\ 2011; Levshakov \etal\ 2011).
Recent detections of methanol (CH$_3$OH) in the LMC (Sewi{\l}o \etal\ 2018)
and SMC (Shimonishi \etal\ 2018) make this possibility a very promising option.

Most proposed molecular transitions fall in the frequency range covered by ALMA bands.
It is to expect that future observations with this telescope as well as new
laboratory measurements of the rest frequencies of perspective transitions will help
to reduce significantly the systematic errors and, hence, to answer the question whether
additional coupling due to non-standard fields is indeed present or not.

\section*{Acknowledgments}
We thank the anonymous referee for critical reading of the manuscript.
The work of S.A.L. was partially supported by the Russian Science
Foundation under grant No.~19-12-00157.

%-------------------------Table 1
\begin{table*}
\centering
\caption{The atomic and molecular transitions used in the present study.
Column 2 shows laboratory frequencies and their uncertainties 
taken from the cited in the last column papers.
The corresponding systematic errors in velocity are given in Column 3.
Column 4 lists beamsizes of the observations. 
}
\label{T1}
\begin{tabular}{l r@.l r@.l  r@.l l}
\hline
\multicolumn{1}{c}{Transition} & 
\multicolumn{2}{c}{Frequency} & 
\multicolumn{2}{c}{$\sigma_{\rm sys}$} & 
\multicolumn{2}{c}{$\theta_{\rm obs}$} &  \multicolumn{1}{c}{Reference}  \\
   & \multicolumn{2}{c}{(GHz)} & 
\multicolumn{2}{c}{(km~s$^{-1}$)} & 
\multicolumn{2}{c}{(arcsec)} &  \\    
\hline
\CI\, $^3$P$_2$--$^3$P$_1$ & 809&341970(17) & 0&006 & 26&5 & Haris \& Kramida (2017) \\
CO(7--6)                 & 806&65180600(50) & 0&0019 & 26&5 & Endres \etal\ (2016) \\
\CII\, $^2$P$_{3/2}$--$^2$P$_{1/2}$ & 1900&5369(13) & 0&205 & 12&0   & Cooksy \etal\ (1986) \\
CO(1--0)                      &  115&27120180(50) & 0&0013 & 33&0    & Endres \etal\ (2016) \\
CO(3--2)                      &  345&79598990(50) & 0&0004 & 17&5   & Endres \etal\ (2016) \\          
\hline
\end{tabular}
\end{table*}

%-----------------------------Table 2
\begin{table*}
\centering
\caption{Radial velocities $V_{\scriptscriptstyle\rm LSR}$ 
of the CO rotational lines ($V_{\rm rot}$) and the [C\,{\sc i}], [C\,{\sc ii}]
fine structure lines ($V_{\rm fs}$)
measured in molecular clouds in the LMC and SMC.
$\Delta V$ is the velocity offset between the CO(7--6) and [C\,{\sc i}] lines.
The data were obtained with: 
[C\,{\sc i}], CO(7--6), and [C\,{\sc ii}]~--
the {\it Herschel Space Observatory}, wherein
[C\,{\sc i}] and CO(7--6) were observed simultaneously within the same band; 
CO(1--0)~-- the {\it Australia Telescope National Facility
(ATNF) Mopra Telescope};
CO(3--2)~-- the {\it Atacama Pathfinder Experiment (APEX) Telescope}.
The channel widths used: $(a)$ 0.88 \kms, $(b)$ 1.11 \kms, $(c)$ 0.19 \kms,
$(d)$ 0.79 \kms, $(e)$ 1.00 \kms, $(f)$ 0.17 \kms, $(g)$ 0.69 \kms, 
$(h)$ 0.70 \kms, $(i)$ 1.04 \kms.
The numbers in parentheses correspond to $1\sigma$ statistical errors on the last digits.
}
\label{T2}
\begin{tabular}{l r@.l r@.l  r@.l  r@.l r@.l r@.l}
\hline
\multicolumn{1}{c}{LOSs} & \multicolumn{2}{c}{CO(7--6)} & 
\multicolumn{2}{c}{[C\,{\sc i}] $^3$P$_2$-$^3$P$_1$} & \multicolumn{2}{c}{$\Delta V =$} & 
\multicolumn{2}{c}{CO(1--0)} & \multicolumn{2}{c}{CO(3--2)} & 
\multicolumn{2}{c}{[C\,{\sc ii}] $^2$P$_{3/2}$-$^2$P$_{1/2}$} \\
  & \multicolumn{2}{c}{$V_{\scriptscriptstyle\rm LSR}$,} & 
\multicolumn{2}{c}{$V_{\scriptscriptstyle\rm LSR}$,} & 
\multicolumn{2}{c}{$V_{\rm rot} - V_{\rm fs} $,} &
\multicolumn{2}{c}{$V_{\scriptscriptstyle\rm LSR}$,} & 
\multicolumn{2}{c}{$V_{\scriptscriptstyle\rm LSR}$,} & 
\multicolumn{2}{c}{$V_{\scriptscriptstyle\rm LSR}$,} \\ 
 & \multicolumn{2}{c}{\kms} & \multicolumn{2}{c}{\kms} & \multicolumn{2}{c}{\kms} & \multicolumn{2}{c}{\kms} & 
\multicolumn{2}{c}{\kms} & \multicolumn{2}{c}{\kms} \\
\multicolumn{1}{c}{\tiny (1)} & \multicolumn{2}{c}{\tiny (2)} & \multicolumn{2}{c}{\tiny (3)} &
\multicolumn{2}{c}{\tiny (4)} & \multicolumn{2}{c}{\tiny (5)} & \multicolumn{2}{c}{\tiny (6)} &
\multicolumn{2}{c}{\tiny (7)} \\
\hline
\multicolumn{13}{c}{\it Large Magellanic Clouds}\\

PDR3-NE &  
271&81(2)$^b$ & 271&85(8)$^b$ & $-0$&04(8) & 271&96(6)$^a$ &
272&88(2)$^f$ & 271&85(7)$^d$ \\[-2pt]
 &  271&81(1)$^c$ & 271&86(2)$^c$ & $-0$&05(2) \\

SK-66D35  & 
276&8(4)$^b$ & 276&8(2)$^b$ & 0&0(4) & 276&2(2)$^a$ & 275&90(9)$^e$ &  280&0(3)$^d$ \\

LMC2-NW & 
288&7(7)$^b$ & 288&55(9)$^b$ & 0&2(7) & 288&92(7)$^a$ \\

NT77  &   
217&60(10)$^b$ & 217&9(2)$^b$ & $-0$&3(2) & 218&00(4)$^a$ & 217&90(8)$^f$ &  217&38(5)$^d$ \\

\multicolumn{13}{c}{\it Small Magellanic Clouds}\\

SMC-NE-3g & 
169&3(3)$^b$ & 169&1(3)$^b$ & 0&2(4) & 168&99(4)$^h$ & 169&2(3)$^i$ &  169&1(3)$^d$ \\

SMC-NE-1a &
149&18(16)$^b$ & 149&30(16)$^b$ & $-$0&1(2) & 148&88(4)$^h$ & 148&8(6)$^g$ &  148&93(13)$^d$ \\

SMC-B2-6  & 
120&8(2)$^b$ & 120&56(13)$^b$ & 0&2(2) & 120&55(5)$^a$ & 120&1(2)$^g$ & 121&01(8)$^d$ \\

SMC-LIRS36 & 
126&29(6)$^b$ & 126&30(7)$^b$ & $-0$&01(10) & 126&29(4)$^a$ & 126&41(8)$^g$ &  126&49(7)$^d$ \\

SMC-LIRS49 & 
114&4(2)$^b$ & 114&68(5)$^b$ & $-0$&3(2) & 114&44(19)$^a$ & 114&50(17)$^g$ & 114&98(16)$^d$ \\
\hline
\end{tabular}
\end{table*}

%-------------------------------------------Figures

% Don't change these lines
\bsp    % typesetting comment
\label{lastpage}
\end{document}